\documentstyle{l-aa}
\input{psfig}
\begin{document}
\thesaurus{05(05.01.1;10.15.2 NGC 1662)}
\title{Proper motion and membership determination in the young
open cluster NGC1662} 
\author{W. S. Dias, R. Boczko, J. I. B. Camargo, R. Teixeira 
and P. Benevides-Soares}
\offprints{W. S. Dias}
\institute{
Universidade de S\~ao Paulo, Dept. de Astronomia,  CP 3386, 
S\~ao Paulo 01060-970, Brazil\\
 email: wilton@iagusp.usp.br}

\date{Received ; accepted 13.03.2000}
  
\maketitle
\markboth{Dias, W. et al. - Proper motions in NGC 1662}{}

\begin{abstract}

Relative proper motions and cluster membership probabilities of 30 stars
within an area of $14^{\prime}$ of diameter centered in the open cluster NGC1662 
($\alpha=04^{\rm h}48^{\rm m}$, 
$\delta=+10^{\rm o}56^{\prime}$) are determined by combining positions of stars observed by the Valinhos CCD meridian
circle with those from other astrometric catalogues.

Our basic source for first epoch of stars brighter than B$=$13 magnitudes is the AC2000 Catalogue,
which provides a time baseline of about 90 years, whereas the USNO--A2.0 Catalogue is 
used for fainter stars, providing a time baseline of about 40 years 
in this region.

Average accuracies in proper motion of 2 mas/yr and 7 mas/yr are achieved,
respectively, when the AC2000 or the USNO--A2.0 is used.

Membership determination is obtained by applying the Zhao \& He (\cite{zha}) method. The cluster proper motion was
found to be ($-3.2\pm 0.6$) mas/yr in right ascension and ($-1.6\pm 0.6$) mas/yr in 
declination.

\end{abstract}

\section{Introduction}
 Open clusters are of special interest since their investigation 
may lead to a deeper insight in kinematics and star formation in the Galaxy, 
as well as contribute to the 
understanding of the Galactic structure. 
Some hundreds of such clusters are known and recent 
contributions indicate an increasing interest in proper motion determination.

In fact, the knowledge of the tangential component of the movement 
plays an important role when membership criteria through astrometrical techniques
are concerned, since a common trend is the only characteristic to be searched for. 
Moreover, in some few cases, the proper motions are accurate enough to allow a study of the 
internal velocity dispersion, providing additional data on the cluster dynamics.

NGC1662 ($\alpha=04^{\rm h}48^{\rm m}27^{\rm s}$, 
$\delta=+10^{\rm o}56^{\prime}13^{\prime\prime}$,  J2000.0) is a well known open cluster.
It is situated in the direction of the anticenter ($l=188^{\rm o}$) and relatively distant from the Galactic plane
($b=20,5^{\rm o}$), corresponding to Z=140 pc for a distance of 378 pc (Hassan \cite{hass}). This distance to the 
plane  may seem surprising for such a young cluster (logt$=$8.11 years) (Hassan \cite{hass}) and suggests 
that the formation mechanism could have been the collision of a high velocity cloud with the gas of the 
Galactic plane (L\'epine \& Duvert \cite{lep}).
 
Many of its properties
are available in the literature such as angular diameter (12$^{\prime}$) (Lyng\aa \hskip 4pt \cite{lyn});
metallicity ($-$0.232) and colour excess (0.34) (Cameron \cite{cam}). Notwithstanding its large content
of bright stars (V$\leq$13.0 mags), no kinematic study was found for this cluster.

In this paper, we present the results of an analysis of membership obtained through proper motions.
The velocity of the centroid of the cluster is also obtained in the process.

A detailed description
of the employed techniques, of the observational material, and an account of the
involved astrometric catalogues, is provided along with an analysis of the final 
results.

\section{Observational data}

Since 1996, an observational program for open clusters using the Valinhos CCD meridian circle (hereinafter VMC) (Viateau 
 et al. \cite{via}) is under way, 
with the aim of proper motion determination.

In this work, only observations performed during the year 1996 were used. The observational 
procedure is entirely differential, and data reduction was made with respect to HIPPARCOS system as materialized by TYCHO catalogue
(ESA \cite{esa}). 
Reference stars within the limits of the cluster were deleted from our 
reference positions file, so that members and candidates could have the same treatment 
during the reduction process.

\begin{table}[hhh]
\caption[]{Astrometric characteristics of Valinhos observational data in the strip centered in NGC1662} 
\begin{flushleft}
\begin{tabular} {lc}
\hline
\hline
Number of stars                   &  400      \\
Strip lenght and width            & 40$^{\rm m}$ and $14^{\prime}$\\
Mean precision in right ascension & $0''.08$  \\  
Mean precision in declination     &  $0''.09$ \\ 
Mean epoch of observations        & 1996.6    \\
Limiting magnitude                & 15        \\
Reference system                  & HIPPARCOS \\
Equinox                           & J2000.0   \\
Reference catalogue                 & Tycho     \\
\hline
\hline
\end{tabular}
\end{flushleft}
\end{table}

\begin{table*}
\caption[]{Catalogues's characteristics in the zone of the cluster NGC1662. For each catalogue, column 2 gives the positional standard
error, column 3 gives the reference system, column 4 gives the number of stars within the zone, column 5 gives the mean epoch 
of observation and column 6 gives the magnitude limit.}
\begin{flushleft}
\begin{tabular}{lcccccc}
\hline
\hline
Catalog    & Mean Precision  & System     & n stars & $\rm {Mean \;Epoch}\over{1900.+}$ & Magnitude Upper Limit\\  
\hline
USNO--A2.0 & $0''.25$        & Hipparcos  & 8122  & 55         &     B and R $\leq$ 20\\ 
TAC        & $0''.09$        & Hipparcos  & 62    & 81         &           B $\leq$ 12\\ 
AC2000     & $0''.30$        & Hipparcos  & 348   & 07         &           B $\leq$ 13\\ 
Tycho      & $0''.03$        & Hipparcos  & 18    & 91         &           V $\leq$ 13\\
PPM North  & $0''.30$        & FK5        & 31    & 31         &           V $\leq$ 11\\ 
SAO        & $0''.17$        & FK5        & 29    & 40         &\hskip 3pt V $\leq$ 9.5\\  
\hline
\hline
\end{tabular}
\end{flushleft}
\end{table*}

The resulting data is, therefore, a catalogue of positions for stars brighter
than V$=$15.0 mags. 

The main astrometric characteristics of VMC observational data for the cluster field are given in Table 1 (Viateau 
et al. \cite{via}).

The large range of right ascension was necessary to assure the observation of enough
reference stars.

\begin{table*}
\caption[]{Positions and proper motions of stars in the region of the cluster NGC 1662. 
The coordinates are given for equinox J2000.0 and mean epoch of observation 1996.6. $\sigma$ denotes the standard deviation.
P is the estimated
membership probability. In the last column are given the catalogues used to determine the proper motions of the stars: 
A=AC2000; P=PPM; S=SAO, T=TAC; U=USNO; Ty=Tycho. The numbers with `*' represent the sequence number in the appropriate zone of 
the USNO catalogue.}
\begin{flushleft}
\begin{tabular}{lccccccccccc}
\hline
\hline
Tycho    & $\alpha$ (J2000)   & $\delta$ (J2000)  & mag  & $\sigma_{\alpha}$& $\sigma_{\delta}$ &$\mu_{\alpha}$  & $\mu_{\delta}$ & $\sigma_{\mu{\alpha}}$& $\sigma_{\mu{\delta}}$& P\\ 
id.   &  hh mm ss          & $^{\rm o}\;\;^{\prime}\;\;^{\prime\prime}$       &  V    & (mas)             &  (mas)              &  (mas/yr)             & (mas/yr)      &  (mas/yr)             & (mas/yr)\\ 
\hline
 687   469\hskip 4pt  1 &4 48 00.572   &10 57 10.57 &10.23   &74   &63    &-4.5              &\hskip 4pt -2.5     &\hskip 4pt 0.8     &\hskip 4pt 0.6  &0.99 &ATTy\\
   3081 *       &4 48 01.140   &10 52 26.63 &13.90   &206  &289   &\hskip 4pt 8.1    &-16.9               &\hskip 4pt 8.9     &10.2 &0.66 &U\\
   3084 *       &4 48 02.478   &10 57 09.34 &14.58   &34   &295   &\hskip 4pt 2.2    &\hskip 4pt -9.8     &\hskip 4pt 7.4     &10.3 &0.58 &U\\
   230047       &4 48 04.947   &10 59 21.19 &11.92   &73   &101   &-0.3              &\hskip 8pt  0.9     &\hskip 4pt 1.2     &\hskip 4pt 1.4  &0.60 &A\\
 687   472\hskip 4pt  1  &4 48 15.119   &11 00 33.75 &11.41   &49   &58    &-4.0              &\hskip 4pt -0.6     &\hskip 4pt 0.6     &\hskip 4pt 0.8  &0.99 &ATy\\
 687   277\hskip 4pt  1  &4 48 15.987   &11 01 04.43 &09.11   &69   &62    &\hskip 4pt 0.1    &\hskip 8pt  4.8     &\hskip 4pt 0.9     &\hskip 4pt 1.1  &0.00 &APTTy\\
   3088 *       &4 48 21.615   &10 54 19.72 &14.32   &87   &90    &-2.2              &\hskip 4pt -9.8     &\hskip 4pt 7.6     &\hskip 4pt 7.7  &0.63 &U\\
   230095       &4 48 22.936   &10 53 13.96 &11.78   &83   &145   &-2.1              &\hskip 4pt -2.4     &\hskip 4pt 1.1     &\hskip 4pt 1.7  &0.91 &A\\
 687   583\hskip 4pt  1  &4 48 24.937   &10 55 09.04 &09.58   &40   &101   &\hskip 4pt 0.7    &\hskip 4pt -0.1     &\hskip 4pt 0.8     &\hskip 4pt 0.5  &0.02 &APSTTy\\
 687 \hskip 4pt 61\hskip 4pt  1  &4 48 26.386   &11 00 42.60 &10.94   &24   &51    &-3.7              &\hskip 4pt -1.2     &\hskip 4pt 0.9     &\hskip 4pt 1.0  &0.98 &ATTy\\
   230104       &4 48 27.311   &10 55 14.43 &11.28   &132  &31    &\hskip 4pt 5.1    &\hskip 4pt -5.5     &\hskip 4pt 1.6     &\hskip 4pt 0.5  &0.00 &A\\
 687   580\hskip 4pt1  &4 48 27.842   &10 55 08.31 &09.91   &22   &75    &-1.7              &\hskip 8pt  0.8     &\hskip 4pt 0.9     &\hskip 4pt 0.8  &0.76 &APTy\\
 687   481\hskip 4pt1  &4 48 27.872   &10 55 45.60 &10.54   &14   &60    &\hskip 4pt 0.3    &\hskip 4pt -1.2     &\hskip 4pt 0.9     &\hskip 4pt 0.7  &0.27 &APTTy\\
   230109       &4 48 29.051   &11 00 11.64 &11.72   &35   &66    &-2.8              &\hskip 4pt -2.3     &\hskip 4pt 1.3     &\hskip 4pt 1.3  &0.95 &A\\
 687   496\hskip 4pt1  &4 48 29.505   &10 55 48.14 &08.42   &56   &60    &\hskip 4pt 9.9    &\hskip 4pt -0.7     &\hskip 4pt 1.3     &\hskip 4pt 1.1  &0.00 &PSTTy\\
   3239 *       &4 48 30.727   &10 59 02.00 &14.40   &174  &369   &-4.0              &-13.0               &\hskip 4pt 8.5     &11.6 &0.64 &U\\
 687   739\hskip 4pt1  &4 48 32.079   &10 57 58.96 &08.71   &38   &70    &\hskip 4pt 4.4    &\hskip 8pt  1.1     &\hskip 4pt 1.2     &\hskip 4pt 1.2  &0.00 &PSTTy\\
 687   556\hskip 4pt1  &4 48 34.488   &10 56 43.98 &09.03   &32   &77    &-2.0              &\hskip 4pt -1.1     &\hskip 4pt 0.6     &\hskip 4pt 0.6  &0.96 &APSTTy\\
  3260 *       &4 48 35.437   &11 00 16.72 &13.89   &154  &269   &-5.5              &-14.7               &\hskip 4pt 8.2     &\hskip 4pt 9.9 &0.71 &U\\
  467995       &4 48 37.533   &10 55 33.52 &12.68   &70   &285   &-1.4              &\hskip 4pt -4.3     &\hskip 4pt 3.7     &\hskip 4pt 4.8  &0.66 &A\\
   3283 *       &4 48 39.896   &10 53 33.32 &14.86   &21   &355   &\hskip 4pt 7.7    &-12.0               &\hskip 4pt 7.4     &11.4 &0.62 &U\\
 687   405\hskip 4pt1  &4 48 41.391   &10 52 02.21 &09.41   &55   &69    &-3.6              &\hskip 4pt -2.8     &\hskip 4pt 0.3     &\hskip 4pt 0.5  &0.99 &APTTy\\
 687   664\hskip 4pt1  &4 48 42.208   &10 59 33.27 &10.84   &50   &36    &-4.2              &\hskip 4pt -1.2     &\hskip 4pt 0.4     &\hskip 4pt 0.5  &0.99 &ATTy\\
   230145       &4 48 43.306   &10 50 31.87 &12.56   &53   &87    &-3.8              &\hskip 4pt -3.1     &\hskip 4pt 1.5     &\hskip 4pt 1.4  &0.96 &A\\
 687   555\hskip 4pt1  &4 48 44.426   &10 53 21.32 &09.47   &60   &26    &\hskip 4pt 0.8    &\hskip 4pt -4.0     &\hskip 4pt 0.8     &\hskip 4pt 0.7  &0.01 &APTTy\\
  3330 *       &4 48 50.033   &10 57 33.92 &14.28   &351  &221   &-7.7              &\hskip 4pt -9.1     &11.3               &\hskip 4pt 9.1  &0.67 &U\\
   3359 *       &4 48 56.021   &10 55 47.40 &14.90   &240  &95    &\hskip 4pt 2.9    &\hskip 4pt -8.6     &\hskip 4pt 9.4     &\hskip 4pt 7.7  &0.52 &U\\
   3373 *       &4 48 58.183   &10 54 33.25 &14.39   &11   &237   &\hskip 4pt 4.8    &\hskip 4pt -7.3     &\hskip 4pt 7.3     &\hskip 4pt 9.4  &0.53 &U\\
 687  1284\hskip 4pt1  &4 49 00.580   &10 50 33.04 &11.75   &75   &47    &-3.3              &\hskip 8pt  6.0     &\hskip 4pt 3.7     &\hskip 4pt 3.4  &0.77 &ATy\\
\hline
\end{tabular}
\end{flushleft}
\end{table*}

It should be pointed out that our sample is not complete, in the sense of the cluster membership, for stars with magnitude
 V$>$15. In fact, 
the star density of USNO-A2.0 (Monet { et al.} \cite{dave}) corresponds to more than 10 times the number of detectable 
stars by VMC,  which is
 a strong indication 
that many of the cluster members may have not been considered in this study. 

The observed star numbers in the region of the cluster are listed in 
Table 3. 
The  first column contains Tycho numbers, when applicable. 
Columns two and three give the J2000 star equatorial coordinates at 
 mean epoch of observation, 1996.6.
Column four lists the visual magnitudes. 

The data and a complete cross identification are available on request.

\section{First epoch catalogues}

Individual proper motions were determined by combining current epoch observations,
as obtained from the VMC, with those provided by previous astrometric catalogues 
(see Table 2).

Our main sources for faint (V$>$13 magnitudes) and bright (V$\leq$13 magnitudes)
stars were the USNO--A2.0 and the AC2000 
(Urban et al. \cite{urbb}a) catalogues, respectively.
In addition to them, TAC (Zacharias et al. \cite{nz}), TYCHO, PPM (R\"oser and Bastian \cite{roe}, Bastian \cite{bas})  and 
SAO (SAO \cite{sao})  catalogues were used so that proper motions could be enhanced for bright stars. Since the last two catalogues 
are given in the FK5 system, some procedure to change their positions to HIPPARCOS system was necessary to compute the proper 
motion.

In view of the small extent of the observed field, the reference system transfer was accomplished simply by averaging the 
differences of the common PPM/SAO and HIPPARCOS stars.

\section{Proper Motions}

The first step in  proper motion determination is the identification of the object  in view in the available catalogues. 

\begin{figure}[ht]
\centerline{\psfig{file=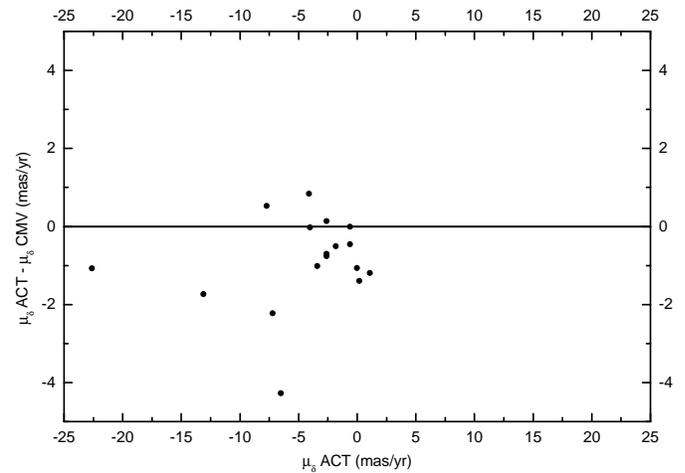,angle=0,width=9.0truecm}}
\vspace{0cm}
\caption[ ]{Comparision of proper motions with ACT catalog in $\mu_{\delta}$}
\label{aaa}
\end{figure}

\begin{figure}[ht]
\centerline{\psfig{file=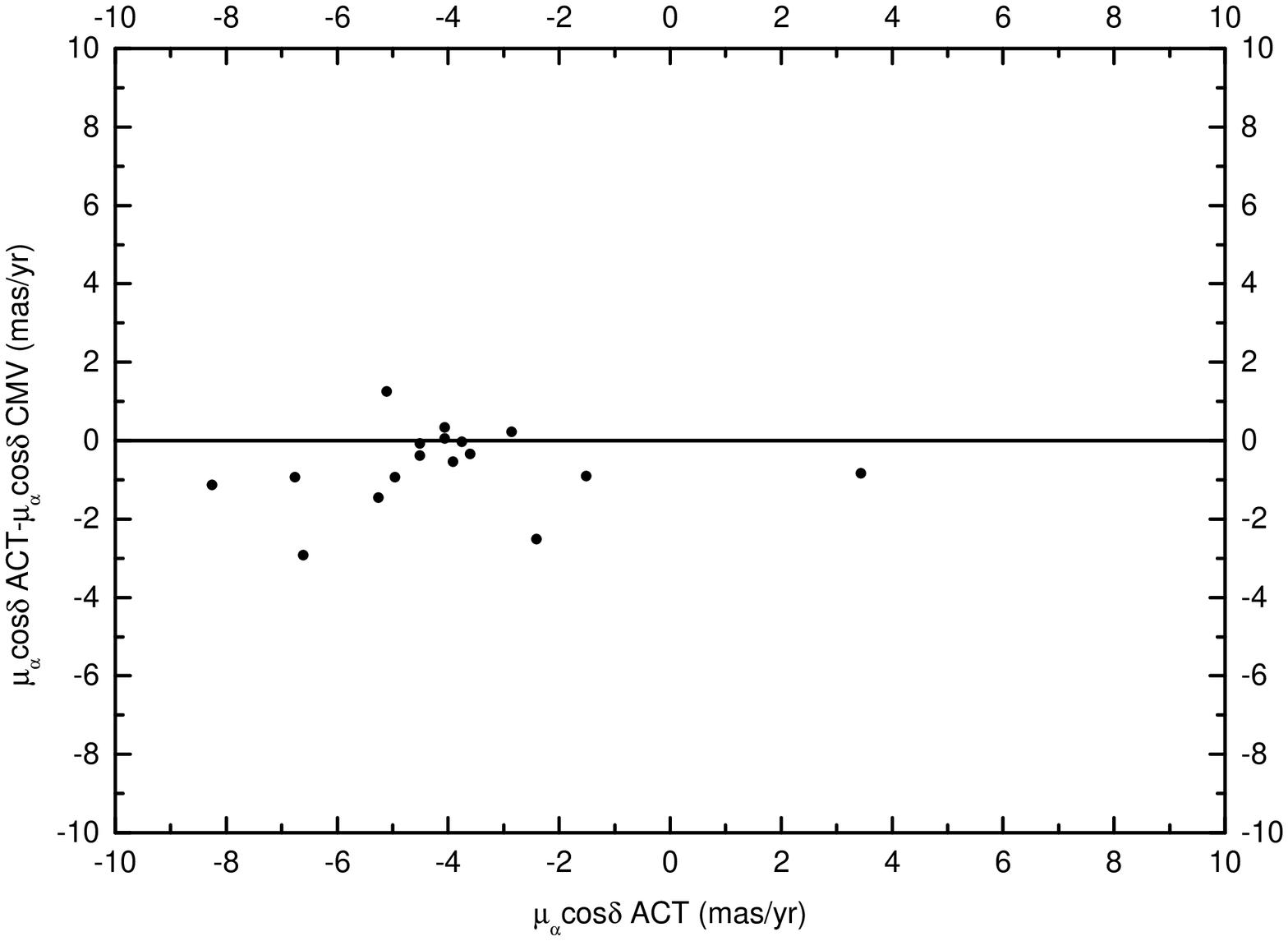,angle=0,width=9.0 cm}}
\vspace{0cm}
\caption[ ]{Comparision of proper motions with ACT catalog in $\mu_{\alpha}\cos\delta$}
\label{aaa}
\end{figure}

Next, we carried out a standard linear weighted least squares solution for the proper motion,
with the weights taken from each individual catalogue documentation.

\subsection{Comparison with ACT proper motions}

To check the quality of our proper motions, we determined the mean square differences for 17 common objects with ACT catalogue 
(Urban et al. \cite{urba}b) obtaining 1.4 mas/yr  and  2.1 mas/yr in $\mu_{\delta}$ and $\mu_{\alpha}\cos\delta$. 

In spite of the near equality between our proper motions and those of ACT (Figs. 1 and 2), we must 
refrain from making definitive conclusions,
since the first epoch for both catalogues is the same, so that strong correlation between both sets is likely. 

\section{Membership Determination}

Usually, the most accurate way to distinguish cluster members from field stars is based on kinematic data, specially on proper motions 
obtained with long time interval between measured positions ( van Leeuwen \cite{lee}).

To determine the membership probabilities of each observed star we adopted the Vasilevskis \& Rach (\cite{vas}) method to provide initial values
of the parameters to be used in Sanders (\cite{ande}), Uribe \& Brieva (\cite{uri}) and Zhao \& He (\cite{zha}) methods.

According to the method suggested by Zhao et al. (\cite{zzz}), proper motions different from the mean by more than four times the 
standard deviation were discarded. 

As expected, the results obtained by all methods were very similar. We consider that the finest results were those obtained by
means of the Zhao \& He method, since heterogeneity can be accommodated to the data accuracy.

In Table 4 we can see the distribution parameters determined by the Zhao \& He (\cite{zha}) method. The meaning of the symbols in Table 4 are as follows:  
$N_{c}$ is the cluster stellar contents; $N_{f}$ is the number of field stars;
($\mu_{xc}$,$\mu_{yc}$)  are the cluster proper motions in x and y; ($\mu_{xf}$,$\mu_{yf}$) are the mean proper motions of field stars in x and y; 
($\sigma_{xf}$,$\sigma_{yf}$) are the dispersions in x and y of the field stars proper motions; $\sigma_{c}$ is the dispersion of cluster stars 
proper motions and $\theta$ is the orientation angle of the minor axis of elliptical field stars proper motions distribution.

With the frequency function parameters we could determine the individual probability of the membership of each star
in the cluster, as suggested by Zhao \& He (\cite{zha}).

In Table 3 we give the kinematic parameters of stars observed by the VMC in the cluster field. 

\begin{table}
\caption[]{Parameters obtained from Zhao and He method to NGC 1662.}
\begin{flushleft}
\begin{tabular}{ccccc}
\hline
\hline
 $N_{c}$ & $\mu_{xc}$ (mas/yr) & $\mu_{yc}$ (mas/yr)& $\sigma_{c}$   \\
   20             &  -3.2         & -1.6               & 0.6          \\
\hline
$N_{f}$ &  $\mu_{xf}$ (mas/yr)& $\mu_{yf}$ (mas/yr)&$\sigma_{xf}$ & $\sigma_{yf}$\\
  10    &    2.4              & -1.2                &  3.8          & 4.0 \\
\hline
 $\theta$\\
 $72.4^{\rm o}$\\
\hline
\hline 
\end{tabular}
\end{flushleft}
\end{table}

\section{Conclusions}

The determination of proper motions by means of combining present epoch observations with ancient positions, 
is a simply obtainable and valuable source of information on stellar kinematics.

In this paper we have combined our meridian circle observations with the ninety years old AC2000 positions and other
data to asserting the membership of 30 stars in the NGC1662 neighbourhood. The results are that 20 of these stars 
have high membership probabilities. We have obtained the overall cluster proper motion, namely, $\mu_{l}= -3.6$ mas/yr and
 $\mu_{b}= 0.6$ mas/yr, in galactic coordinates. 

\begin{acknowledgements}

We would like to thank Dr. J. L\'epine, Dr. N. Zacharias and Dr. S. Urban for their suggestions and we are specially grateful
to Mr. W. Monteiro for his collaboration in the observations at Valinhos, as well
as all those who contributed to this work. W. S. Dias is supported by CNPq. 
\end{acknowledgements}

\end{document}